\documentclass[a4paper,showkeys,floatfix,aps,pre,reprint,groupedaddress]{revtex4-1}
\usepackage{amsmath,amssymb}
\usepackage{graphics,graphicx}
\usepackage{dcolumn,bm}
\usepackage{psfrag}
\usepackage{enumitem}

\begin{document}

\title{Stability in fiber bundle model : Existence of strong links and the effect of disorder}

\author{Subhadeep Roy}
\email{sroy@eri.u-tokyo.ac.jp}
\affiliation{Earthquake Research Institute, University of Tokyo, 1-1-1 Yayoi, Bunkyo, 113-0032 Tokyo, Japan.}

\date{\today}

\begin{abstract}
In this paper I have studied the fiber bundle model with a fraction $\alpha$ of infinitely strong fibers. Inclusion of such unbreakable fraction has been proven to affect the failure process in early studies, especially around a critical value $\alpha_c$. The present work has a twofold purpose: (i) study of failure abruptness, mainly the brittle to quasi-brittle transition point ($\delta_c$) with varying $\alpha$ and (ii) variation of $\alpha_c$ as we change the disorder introduced in the model. The brittle to quasi-brittle transition is confirmed from the failure abruptness. On the other hand, the $\alpha_c$ is obtained from the knowledge of failure abruptness and the statistics of avalanches. It is observed that $\delta_c$ scales to lower values, suggesting more quasi-brittle like continuous failure when $\alpha$ is increased. Also, the critical fraction $\alpha_c$, required to make the model deviate from the conventional results, increases with decreasing strength of disorder. The analytical expression for $\alpha_c$ shows good agreement with the numerical result. Finally, the findings in the paper are compared with previous results as well as with the real life application of composite materials.   
\end{abstract}


\maketitle


\section{Introduction}
Existence of a strong link plays a crucial role while studying the stability during the failure process in fiber bundle model \cite{Duxbury1,RMP,Fiber1}. Recent studies show that the existence of an unbreakable fraction affects the burst statistics in global load sharing \cite{Hidalgo} as well as local load sharing scheme \cite{Kovacs}. The average size of maximum burst shows an abrupt change around a critical fraction $\alpha_c$ of strong links. Also, recently it was observed that introduction of disorder greatly affects the failure process in the model \cite{sroyepl}. In that study, a critical strength of disorder was observed that separates brittle like abrupt failure from non-abrupt quasi-brittle like failure. Such critical disorder strength is function of system size \cite{sroy} in the local load sharing scheme. Though the disorder plays significant role in LLS model, the uniqueness of such critical disorder is lost due to the system size effect \cite{sroy,Gomez,Pradhan1} present there. For the present work, I have concentrated only in the mean field limit, for both analytical as well as numerical studies.   

The present work has two main purposes that revolves around the idea that a disordered system with a variable strength of disorder contains a fraction of infinitely strong links. 
\begin{enumerate}[label=(\roman*)]
\item The first purpose of the paper will be studying how the existence of strong links affects the failure process, especially the failure abruptness. The study mainly includes the behavior of brittle to quasi-brittle transition point, which was already observed in the mean field limit \cite{sroyepl,Chandryee}, as we treat the fraction $\alpha$ of strong links as the variable. Two extreme limits of this variable can be understood: (a) $\alpha=0$, which is the conventional limit and (b) $\alpha=1$, where each and every fiber has infinite strength and the model does not evolve at all. The present study discusses the nature of failure in all possible $\alpha$ values.  
\item The other part of the paper is dedicated to understand the response of $\alpha_c$ as the disorder is varied. A previous study \cite{Hidalgo} in the mean field limit claims that for uniform threshold distribution [0,1] (introducing a constant disorder), half of the bundle should have infinite strength to change the conventional avalanche behavior: scale free decay with an universal exponent $-5/2$ \cite{Hemmer,Hansen,Kloster}. Here I have studied how $\alpha_c$ changes when the strength of disorder is continuously varied.  
\end{enumerate} 

The idea of introducing some infinitely strong fiber in the conventional bundle can be connected loosely with the manufacture of composite materials \cite{Composite1,Dieter,Lawn,Lee,Mallick}. In material science, composites are prepared by mixing two materials of different properties in certain proportion that usually comes with higher strength \cite{Chermoshentseva} and toughness \cite{Zhao} than the component materials. Fiber reinforced composite \cite{FibReinforce,harlow78,harlow91,phoenix74,phoenix75,Curtin1,Curtin2} is a good example in such a case where fibers with high strength are embedded with a carrier matrix. There are some recent works \cite{Hader,Hader1,Raischel} in fiber bundle model in comparison with composite materials. The present work shows how the strength and failure abruptness of the composite FBM, prepared by mixing a fraction of infinitely strong fibers in the conventional model, is affected in presence of variable disorder.

In the next section a description of the model is provided, followed by the analytical results in the mean field limit (Section III). Section IV is dedicated to the numerical results performed with $10^5$ fibers and a large set ($\approx 10^4$) of configurations. Finally, in Section V I have provided the discussions and concluding remarks on the work.  


\section{Description of the model}
After its introduction by Pierce in 1926 \cite{Pierce}, fiber bundle model has been proven to be important yet arguable the simplest model to study failure process. A conventional fiber bundle model consists of fibers or Hookean springs, attached between two parallel plates. The plates are pulled apart by a force $F$, creating a stress $\sigma=F/L$ on $L$ fibers. Once the stress crosses the breaking threshold of a particular fiber, chosen from a random distribution, that fiber breaks irreversibly. The stress of a broken fibers is then redistributed either globally among all surviving fibers (global load sharing or GLS scheme) or among the surviving nearest neighbors only (local load sharing or LLS scheme). For the GLS scheme \cite{Pierce, Daniels} no stress concentration occurs anywhere around the failed fibers as the stress of the failed fibers is shared among all surviving fibers democratically. On the other hand, in LLS scheme \cite{Phoenix,Smith,Newman,Harlow3,Smith2}, stress concentration is observed near a broken patch (series of broken fibers) and increases with the length of such patches. After such redistribution, the load per fiber increases initiating failure of more fibers and starting an avalanche. At the end of such avalanche either all fibers are broken (suggesting global failure) or the bundle comes to a stable state with few broken fibers where an increment of external stress is required to make the model evolve further. The last applied stress just before global failure is considered to be the nominal stress or strength of the bundle. 

In this work the conventional model is modified by considering a fraction of total fibers to be infinitely strong and therefore can bear any amount of stress without breaking. This kind of work is already studied in fiber bundle model with both GLS \cite{Hidalgo} and LLS \cite{Kovacs} schemes. We have carried out the study with varying disorder in the mean filed or GLS limit and observe how above findings are affected when the strength of disorder is varied.

If there are initially $L$ fibers in the model then among them lets assume $\alpha$ fraction ($\alpha L$ number of fibers) is unbreakable and does not contribute to the evolution of the model. A certain amount of applied stress breaks some fibers among the rest ($1-\alpha$) fraction and increases stress per fiber that leads to avalanches. The infinitely strong fibers carry the extra stress, due to redistribution, without breaking and does not contribute to the avalanche process. So the dynamics of the model is mainly determined by $(1-\alpha)L$ conventional fibers that have  random but finite breaking thresholds. The existence of such $\alpha$ fraction makes the model evolve slowly as number of fibers broken at a particular stress decrease that lowers the local stress profile than the conventional model.

The next section contains some analytical results for the model, dealing with the variation of $\alpha_c$ with disorder $\delta$ as well as the behavior of brittle to quasi-brittle transition point with varying $\alpha$ values. 
 

\section{Analytical approach}
For analytical calculation let us assume that a fraction of fibers $\alpha$ in the model is too strong to break. A stress $\sigma_0$ is applied externally creating a stress per fiber $\sigma$. We have shown the analytical calculations for uniform distribution of threshold stress. Also, a different case with power law threshold distribution is adopted in order to approach high disorder limit. This will be discussed later in this paper. 


\subsection{Critical fraction for strong links}
In case of a particular distribution $P(\sigma)$ of threshold strength values, we can relate the externally applied stress ($\sigma_0$) with local stress per fiber ($\sigma$) as : 
\begin{equation}
\label{eq:1}
\sigma_0=(1-\alpha)\big[1-P(\sigma)\big]\sigma+\alpha\sigma 
\end{equation}
The second part of Eq.\ref{eq:1} shows the stress carried by the infinitely strong fibers while the first part gives the stress carried by conventional fibers after certain redistribution (depending on $\sigma_0$).

For an uniform distribution of width $2\delta$ and mean at $0.5$ we get $P(\sigma)=\displaystyle\frac{\sigma-a}{2\delta}$, $\delta$ being the strength of disorder and $a (=0.5-\delta)$ is the minimum of the threshold distribution . For that particular case Eq. \ref{eq:1} can be written as 
\begin{equation}
\label{eq:2}
\sigma_0=(1-\alpha)\left[1-\displaystyle\frac{(\sigma-a)}{2\delta}\right]\sigma+\alpha\sigma 
\end{equation}
This will give a parabolic curve at $\alpha=0$. For other $\alpha$ values there will be a curve with a maximum at the unstable point of the model i.e at the critical point of stress per fiber ($\sigma_c$). This point is given by 
\begin{equation}
\label{eq:3}
\displaystyle\frac{d\sigma_0}{d\sigma}\bigg|_{\sigma_0=\sigma_c}=0
\end{equation}
Inserting value $\sigma_0$ from Eq.\ref{eq:2} and using $a=\left(0.5-\delta\right)$ we get 
\begin{eqnarray}
\label{eq:4}
&1-\displaystyle\frac{\sigma_c}{\delta}(1-\alpha)-\displaystyle\frac{0.5-\delta}{2\delta}(1-\alpha)=0 \nonumber \\
&\text{or,} \ \sigma_c(\alpha)=\displaystyle\frac{\delta}{1-\alpha}-\left(\displaystyle\frac{1}{4}-\displaystyle\frac{\delta}{2}\right) \ \ \ 
\end{eqnarray}
Now the maximum value $\sigma$ can attain is $(0.5+\delta)$ which is the maximum of the threshold distribution and after this point we can't get any maximum of the curve. This is the point where $\alpha$ reaches its critical value. So at $\alpha=\alpha_c$ we get $\sigma_c=(0.5+\delta)$. Implying this condition we get this critical fraction of strong links in terms of disorder.
\begin{equation}
\label{eq:5}
\alpha_c=\displaystyle\frac{3-2\delta}{3+2\delta}
\end{equation}
So as we go for higher and higher $\delta$ values, $\alpha_c$ decreases and we need lesser fraction of unbreakable fibers to attain the critical point. 


\subsection{Study of failure abruptness}  
To understand the brittle to quasi-brittle transition point in the model we have to construct the recursion relation for fraction of unbroken bonds. Lets assume that the total $N_u$ number of unbroken fibers are the combination of $N_u^s$ numbers of infinitely strong fibers and $N_u^w$ numbers of conventional fibers. This makes $N_u^s/N_u=\alpha$ and $N_u^w/N_u=1-\alpha$. In recursion relation $N_u^s$ does not have any role. Then, for a uniform distribution with mean at 0.5 and width $2\delta$ the equation of the fraction unbroken can be given by 
\begin{align}
\label{eq:6}
(1-\alpha)-n_u^w&=\displaystyle\int_{a}^{\displaystyle\frac{\sigma_0}{n_u^w+\alpha}}p(\sigma)d\sigma \nonumber \\
                &=\displaystyle\frac{1}{2\delta}\displaystyle\int_{a}^{\displaystyle\frac{\sigma_0}{n_u^w+\alpha}}d\sigma
\end{align}  
Where $n_u^w$ is the fraction of unbroken bonds corresponding to applied stress $\sigma_0$ and $a$ ($=0.5-\delta$) is the minimum of the distribution. Eq.\ref{eq:6} will give a quadratic equation of $n_u^w$ 
\begin{align}
(n_u^w)^2-n_u^w\left(1+\displaystyle\frac{a}{2\delta}-2\alpha\right)+\left(\alpha^2+\displaystyle\frac{\sigma_0}{2\delta}-\alpha-\displaystyle\frac{a\alpha}{2\delta}\right)=0
\end{align}
The solution to above equation will be 
\begin{eqnarray}
\label{eq:7}
n_u^w= \displaystyle\frac{1}{2}\Bigg[\left(1+\displaystyle\frac{a}{2\delta}-2\alpha\right) \pm \nonumber \\  
\sqrt{\left(1+\displaystyle\frac{a}{2\delta}-2\alpha\right)^2-4\left(\alpha^2+\displaystyle\frac{\sigma_0}{2\delta}-\alpha-\displaystyle\frac{a\alpha}{2\delta}\right)} \Bigg]
\end{eqnarray}
Since at critical point two solutions of Eq.\ref{eq:7} cannot exist, it suggest at critical point the rooted part of above equation will vanish. In that case, the critical fraction unbroken will be given by 
\begin{equation}
\label{eq:9}
n_c=(n_u^w)_c=\displaystyle\frac{1}{2}\left(1+\displaystyle\frac{a}{2\delta}-2\alpha\right)
\end{equation}
For the case $\alpha=0$ above results reduces to conventional results in the model \cite{sroyepl}, where all fibers can break. An abrupt brittle like failure is seen in the model when $n_c=1$. Inserting $n_c=1$ and $a=0.5-\delta_c$ we get the value of $\delta_c$ from above equation as :
\begin{equation}
\label{eq:10}
\delta_c=\displaystyle\frac{1}{2}\displaystyle\frac{1}{(4\alpha+3)}
\end{equation}
For $\alpha=0$ we get $\delta_c=1/6$ which is the exact result we obtained for convention fiber bundle model in the mean field limit \cite{sroyepl}.


\section{Numerical Results}
Numerically, the model has been studied with system size $10^5$ and a large set ($\sim 10^4$) of configurations. Most of the results are generated in the mean field limit with uniform distribution, though the last part of the numerical results are discussed with power law threshold distribution to confirm the universality as well as to approach high disorder limit. Previous numerical studies suggest that there exists a critical fraction $\alpha_c$ of strong links above which the avalanche statistics deviates from the mean field results. The value of $\alpha_c$ is quite high ($=0.5$) \cite{Hidalgo} with GLS scheme and drops to a very low value ($\sim 0.05$) \cite{Kovacs} with local stress concentration. In this paper, I have studied this $\alpha_c$ in details with varying strength of disorder in the mean field limit. Also, the stability of the model during the failure process is discussed with varying $\alpha$ values.
   
Below the numerical results are shown where the threshold values are chosen from a uniform distribution of half width $\delta$ and mean at 0.5. $\delta$ expresses the strength of disorder here.  

\subsection{Stability during failure process}
Fraction of unbroken bond just before the global failure ($n_c$) is proven to be a good measure of failure abruptness in recent studies \cite{sroyepl,sroy,sroyArxiv}. $n_c=1$ corresponds to an abrupt failure as the total model is intact just before the global failure. After the application of a stress, large enough to break the weakest fiber, the bundle becomes unstable and breaks in a single avalanche without any prior warning. $n_c<1$ suggests that the bundle goes through a number of stable states before global failure. At each stable state an increment in applied stress is required. In this section we have discussed the behavior of $n_c$ with varying disorder, for different fraction $\alpha$. As expected, at low disorder, $n_c$ remains at 1.0 and the failure process is brittle like abrupt. On the other hand, at high disorder $n_c<1$ and the model goes through a number of stable states prior to failure point.   
\begin{figure}[ht]
\centering
\includegraphics[width=7cm]{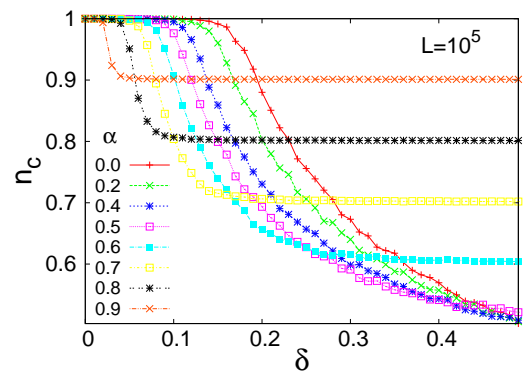}
\caption{Variation of critical fraction unbroken $n_c$ with strength of disorder $\delta$ for different $\alpha$ values. For low $\alpha$, $n_c$ remains at 1 at low disorder strength and gradually decreases to 0.5 with increasing $\delta$. For high $\alpha$, due to the existence of unbreakable fibers, $n_c$ saturates to a certain value ($>0.5$) beyond a certain strength of disorder.}
\label{Failure_Abruptness}
\end{figure}
At low $\alpha$ values, where very small fraction of the fibers are strong, $n_c$ decreases to 0.5 when the disorder reaches $\delta=0.5$. Such behavior remains unchanged in the region $\alpha<0.5$. As we go to high values for $\alpha$, $n_c$ starts saturating after a certain disorder value. The saturation occurs as the remaining fibers are strong enough to bear any stress without breaking. In this section we will mainly concentrate on the point where $n_c$ deviates from 1 and hence from abrupt failure. The disorder at which such deviation takes place is denoted as $\delta_c$. Two extreme limit of the model corresponds to $\alpha=0$ and $\alpha=1$. The former corresponds to the conventional mean field limit where $\delta_c$ is expected to be around $1/6$ \cite{sroyepl} with uniform threshold distribution (mean 0.5 and half-width $\delta$). On the other hand, the later one corresponds to a situation where each and every fiber is unbreakable and the model does not evolve at all. Fig.\ref{Failure_Abruptness} clearly shows that $\delta_c$ approaches lower values as we increase $\alpha$. This in turn, reduces the window of disorder strength within which an abrupt failure is expected. As a result, at high $\alpha$ we will start getting stable states during failure process, even at low strength of disorder.   


\subsection{Strong link and disorder dependence in probability of abrupt failure}
To understand the predictability and stability during failure process, we have studied the probability of abrupt failure with varying disorder and $\alpha$. A recent study \cite{sroyepl} already shows how the predictably is affected by disorder in the mean field limit. Here such behavior is studied with varying fraction of strong links. 
\begin{figure}[ht]
\centering
\includegraphics[width=7cm]{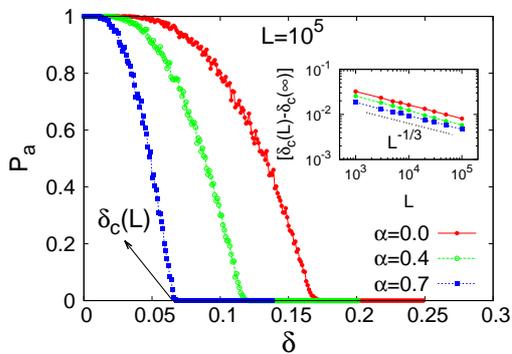}
\caption{$P_a$ vs $\delta$ for $\alpha=0.0$, $0.4$ and $0.7$. For $\delta<\delta_c$, $P_a>0$ and hence there is a non-zero probability of abrupt failure. The inset shows that $\delta_c(L)$ approaches its thermodynamic limit $\delta_c(\infty)$ as: $\delta_c(L)=\delta_c(\infty)+L^{-1/3}$. The scaling remains invariant w.r.t $\alpha$.}
\label{Proability}
\end{figure}
The probability of abrupt failure, $P_a$, is basically defined as the ratio of how many times the model goes through abrupt failure (breaks in a single avalanche) to the total number of observations. Fig.\ref{Proability} shows $P_a$ as a function of disorder strength $\delta$ for different fraction $\alpha$. At low disorder $P_a$ remains at 1 and the failure is abrupt for each and every observation. With increasing $\delta$, $P_a$ gradually decreases to zero. The region $P_a>0$ is denoted as brittle as there exist a non zero probability of abrupt failure. $\delta_c(L)$ is defined as the critical disorder for a particular system size $L$ beyond which $P_a=0$. Fig.\ref{Proability} clearly shows a decreasing $\delta_c(L)$ when $\alpha$ is increased. Also the fall of $P_b$ becomes more and more sharp. This in turn supports our previous claim of decreasing abrupt failure with $\alpha$ values. Also, $\delta_c(L)$ scales down with increasing system size as : $\delta_c(L)=\delta_c(\infty)+L^{-\zeta}$, where $\zeta=1/3$. $\delta_c(\infty)$ is the brittle to quasi-brittle transition point at the thermodynamic limit. The inset shows that the above scaling of $\delta_c(L)$ remains unchanged even when $\alpha$ is varied. 
\begin{figure}[ht]
\centering
\includegraphics[width=7.5cm]{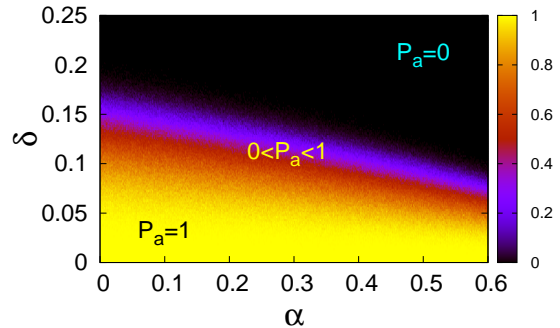}
\caption{$P_a$ as a function of both disorder $\delta$ and fraction of strong link $\alpha$. The yellow and black color corresponds to pure abrupt and pure non-abrupt failure. Within the color gradient, the abruptness in failure process is probabilistic.}
\label{Prob_3d}
\end{figure}

In Fig.\ref{Prob_3d}, the behavior of $P_a$ is discussed while both the parameters $\alpha$ and $\delta$ is varied simultaneously. The color scheme for $P_a$ can be understood as follows: 
\begin{itemize}
\item The yellow color stands for the condition $P_a=1$. The failure process is abrupt here in each and every observation. 
\item The black color corresponds to $P_a=0$ and the failure process is always quasi-brittle like non-abrupt.  
\item The region $0<P_a<1$ is shown the other color gradients. The probability of abrupt failure is variable in this region and decreases with both $\alpha$ and $\delta$. 
\end{itemize}  
Fig.\ref{Prob_3d} shows that at higher $\alpha$, while going from yellow to black region, we cross the color gradient at a lower disorder strength.   


\subsection{Brittle to quasi-brittle transition point}
\begin{figure}[ht]
\centering
\includegraphics[width=6.5cm]{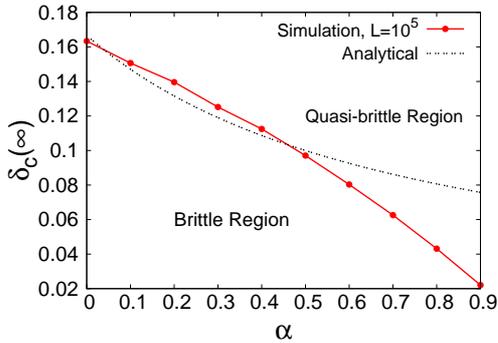}
\caption{Comparison between theoretical and numerical findings on $\delta_c(\infty)$ for different $\alpha$ values. $\delta_c(\infty)$ decreases with increasing $\alpha$ making the quasi-brittle response more and more prominant.}
\label{Deltac}
\end{figure}

We have now reached the point where we can discuss the brittle to quasi-brittle transition point $\delta_c(\infty)$ with continuously varying $\alpha$ values. In Fig.\ref{Deltac} we have shown this numerical variation along with the analytical finding given by Eq.\ref{eq:10}.   
Though the analytical and numerical values of $\delta_c(\infty)$ do not show a good agreement, both agrees to a decreasing behavior of $\delta_c(L)$ with increasing $\alpha$. Above disagreement is quite prominent for higher $\alpha$ values. As per Fig.\ref{Deltac}, $\delta_c(L)$ starts from 1/6 at $\alpha=0$ (the conventional limit) and decreases to 0.02 for $\alpha=0.9$. In this high $\alpha$ limit, the failure process is predictable almost for all strength of disorder.    


\subsection{Strength of the bundle}
\begin{figure}[ht]
\centering
\includegraphics[width=7cm]{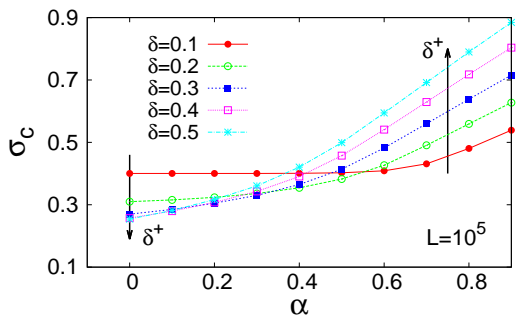}
\caption{Variation of critical strength $\sigma_c$ with $\alpha$ for different strength of disorder $\delta$. $\sigma_c$ increases monotonically with $\alpha$ for any $\delta$ value. The response of $\sigma_c$ against disorder is also modified as $\alpha$ is increased.}
\label{Strength}
\end{figure}

A different way to understand the advantage of introducing a fraction of unbreakable fiber is to monitor how the strength of the bundle is affected by it. Fig.\ref{Strength} shows that the strength of the bundle increases as we include more and more unbreakable fibers. 

We already know that $\alpha=0$ leads to the conventional limit of the model, where with decreasing strength of disorder $\delta$, the failure process becomes more brittle but at the same time strength of the bundle increases. Fig.\ref{Strength} suggests that, as we go to a relatively high $\alpha$ value the response of $\sigma_c$ against $\delta$ reverses and instead of decreasing $\sigma_c$ starts increasing with $\delta$. Combining this finding along with the results of failure abruptness we can conclude that at high $\alpha$ the model operates in an ideal situation where strength of the bundle is high and the failure process is highly predictable due to quasi-brittle like continuous failure.  	 


\subsection{Estimation of $\alpha_c$ from failure abruptness}
\begin{figure}[ht]
\centering
\includegraphics[width=7cm]{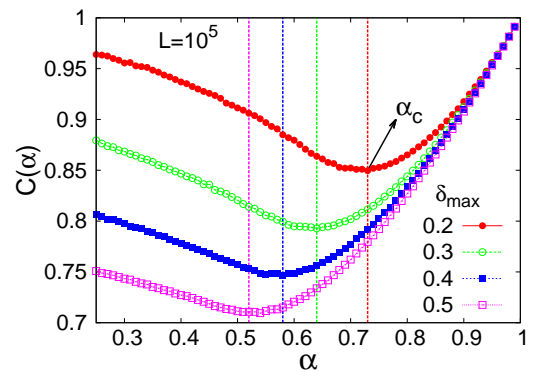}
\caption{$C(\alpha)$ as a function of fraction of the strong link $\alpha$. The minimum of the plot gives us $\alpha_c$, beyond which the strong link plays a crucial role and the model deviates from its conventional results.}
\label{Charesteristic_Function}
\end{figure}

To estimate the critical fraction of strong link we have studied the failure abruptness (Fig.\ref{Failure_Abruptness}) once again. The particular characteristic function we have chosen here is $C(\alpha)$ and will be given by 
\begin{align}
C(\alpha)=\displaystyle\sum_{\delta=0}^{\delta_{max}}n_c(\delta)
\end{align}
Fig.\ref{Charesteristic_Function} shows the variation of $C(\alpha)$ with $\alpha$ for $\delta_{max}$ ranging in between $0.2$ and $0.5$. 
At low $\alpha$ values $n_c$ remains at 1.0 up to high $\delta$ values and then start decreasing until it reaches 0.5 at $\delta=0.5$. $C(\alpha)$ has a high value in that case. The other limit is where $\alpha$ has a high value. In that limit $n_c$ deviates from 1.0 at a very lower $\delta$ but later saturates at a higher value than 0.5, as the rest of the fibers are unbreakable. The characteristic function is also high in this limit. At any intermediate $\alpha$ value $C(\alpha)$ shows a non monotonic behavior. The minimum of the curve corresponds to $\alpha_c$, the critical fraction of strong links, where the model is about to deviate from the conventional limit to the limit where the strong links play crucial role. At $\delta_{max}=0.2$, we obtain the $\alpha_c$ around 0.7. As we increase $\delta_{max}$, the minimum as well as $\alpha_c$ scales down to lower values. Finally, at $\delta_{max}=0.5$, $\alpha_c \approx 0.5$, which matches with the previous claim by Hidalgo et. al \cite{Hidalgo}. Since for $\alpha>\alpha_c$ the strong links are important in determining the failure process, this region can be named as the strong link dominated region.     

The system size effect of $\alpha_c$ is also studied. $\alpha_c$ is observed to show almost no change while system size is increased. The studies are carried out over the range $10^3 \le L \le 10^5$. Over such range of system size, $\alpha_c$ changes by an amount 0.01 roughly. Due to such $L$-independent behavior, it is safe to treat the above $\alpha_c$'s as $\alpha_c(L\rightarrow\infty)$, their values in the thermodynamic limit.     


\subsection{Behavior of maximum burst}
To understand the existence of $\alpha_c$ and its variation with disorder $\delta$, I have also studied how the maximum burst behaves at different conditions. A burst size is defined here as the number of fibers broken in between two consecutive stress increment. The final burst during the failure process has been neglected in above studies. The maximum of the burst are chosen among the rest of the avalanches. $\langle\Delta_{max}\rangle$ is defined as the average over $10^4$ such maximum burst values.  
\begin{figure}[ht]
\centering
\includegraphics[width=7cm]{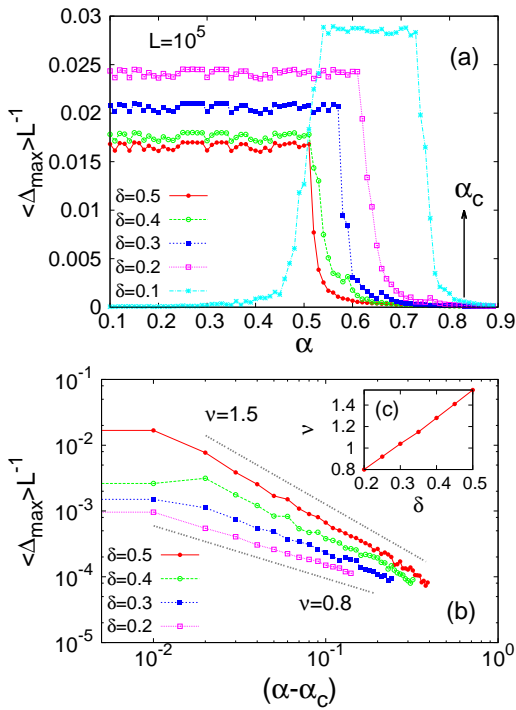}
\caption{(a) Variation of average maximum burst size $\langle\Delta_{max}\rangle$ with $\alpha$. For $\delta>\delta_c$, $\langle\Delta_{max}\rangle$ falls sharply around $\alpha_c$. For $\delta<\delta_c$, $\langle\Delta_{max}\rangle$ shows a non-monotonic behavior but still shows an abrupt change as we go beyond $\alpha_c$. \\ (b) $\langle\Delta_{max}\rangle$ diverges around the critical fraction $\alpha_c$ as: $\langle\Delta_{max}\rangle/L \sim (\alpha-\alpha_c)^{-\nu}$. \\ (c) The inset shows the variation of $\nu$ with disorder.}
\label{Burst_Size_Fall}
\end{figure}
Fig.\ref{Burst_Size_Fall} shows how $\langle\Delta_{max}\rangle$ varies with $\alpha$ at different disorder strength. For $\delta>\delta_c(\infty)$, $\langle\Delta_{max}\rangle$ saturates at a non-zero value for small $\alpha$ (see Fig.\ref{Burst_Size_Fall}(a) for $\delta=0.5$, 0.4, 0.3 and 0.2). As the model crosses $\alpha_c$, $\langle\Delta_{max}\rangle$ shows a sudden decrease and reaches zero gradually. As we decrease $\delta$, this sudden jump in $\langle\Delta_{max}\rangle$ starts taking place at higher $\alpha$, suggesting a shift of $\alpha_c$ towards high values. For $\delta<\delta_c$, $\langle\Delta_{max}\rangle$ shows a non monotonic behavior. $\langle\Delta_{max}\rangle$ remain at a very low value for small $\alpha$; reaches a maximum and falls back again close to $\alpha_c$. Fig.\ref{Burst_Size_Fall}(b) offers a closer look to the divergence of $\langle\Delta_{max}\rangle$ around $\alpha=\alpha_c$. The following scaling is observed:
\begin{align}
\displaystyle\frac{\langle\Delta_{max}\rangle}{L} \sim (\alpha-\alpha_c)^{-\nu} 
\end{align}  
where the exponent $\nu$ has a disorder dependence. As we decrease disorder form $\delta=0.5$, $\nu$ starts to decrease from $1.5$ (matches with earlier claim by Hidalgo et. al \cite{Hidalgo}) and reaches 0.8 at $\delta=0.2$. We have restricted our study to $\delta>\delta_c(\infty)$ as below this disorder  $\alpha_c$ value is quite high and there will be very few points before the model stops evolving. The variation of $\nu$ with $\delta$ is shown in Fig.\ref{Burst_Size_Fall}(c). The increment of $\nu$ is almost linear with strength of disorder $\delta$.  


\subsection{Burst size distribution}
\begin{figure}[ht]
\centering
\includegraphics[width=7cm]{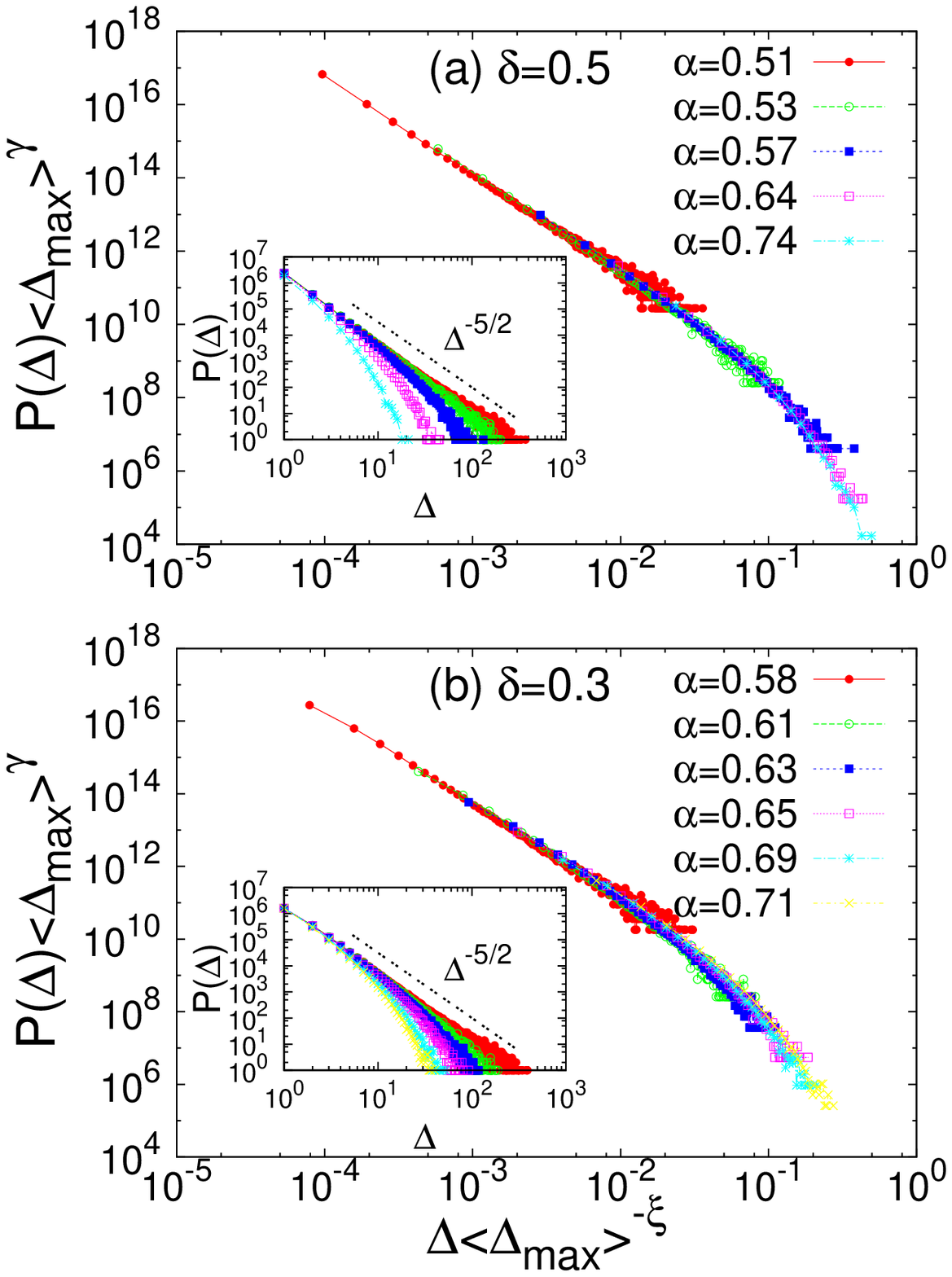}
\caption{Burst size distribution for (a) $\delta=0.5$ and (b) $0.3$. The inset shows the unscaled distribution. We scaled the distribution with following scaling rule: \\ $P(\Delta)=\langle\Delta_{max}\rangle^{-\gamma}\Phi\left(\displaystyle\frac{\Delta}{\langle\Delta_{max}\rangle^\xi}\right)$.}
\label{Avalanche_Size}
\end{figure}

Finally, we have studied the distribution of burst size $\Delta$ and how does it scale at different disorder values. The size of a burst holds the same definition as previous: number of fibers broken in between consecutive stress increment. Fig.\ref{Avalanche_Size} shows the burst size distribution $P(\Delta)$ for $\delta=0.5$ and 0.3.
The figures in the inset show the unscaled burst distribution while in the main figures the following scaling is being adopted: 
\begin{align}\label{scaling}
P(\Delta)=\langle\Delta_{max}\rangle^{-\gamma}\Phi\left(\displaystyle\frac{\Delta}{\langle\Delta_{max}\rangle^\xi}\right)
\end{align}
For $\alpha<\alpha_c$ the model eventually operates similar to the conventional model and the size distribution will be given by: $P(\Delta) \sim \Delta^{-\kappa}$, with $\kappa=-5/2$ \cite{Hemmer,Hansen,Kloster}. As we go beyond $\alpha_c$, $P(\Delta)$ starts deviating from this conventional scale free behavior. For lower disorder such deviation starts at higher $\alpha$ values.  
\begin{table}[ht]
\begin{tabular}{|c|c|c|c|}
\hline
\hspace{0.7cm} $\delta$ \hspace{0.7cm} & \hspace{0.7cm} $\gamma$ \hspace{0.7cm} & \hspace{0.7cm} $\xi$ \hspace{0.7cm} & \hspace{0.3cm} $\gamma/\xi(\approx \kappa)$ \hspace{0.3cm} \\ \hline
0.5 & 3.25 & 1.25 & 2.60 \\ \hline
0.4 & 3.40 & 1.35 & 2.52 \\ \hline
0.3 & 3.45 & 1.38 & 2.50 \\ \hline 
\end{tabular}
\caption{Scaling exponents $\gamma$ and $\xi$ for different $\delta$ is tabulated.}
\label{table:exp}
\end{table}
The scaling exponents $\gamma$ and $\xi$ given by Eq.\ref{scaling} is shown in Table \ref{table:exp} for $\delta=0.5$, 0.4 and 0.3. The exponents at $\delta=0.5$ matches with the previous findings \cite{Hidalgo}. As we decrease $\delta$, both $\gamma$ and $\xi$ increases, satisfying $\gamma=\kappa\xi$ (see Table \ref{table:exp}).


\subsection{Variation of $\alpha_c$ with disorder}
Above behavior of $C(\alpha)$ as well as the study of $\langle\Delta_{max}\rangle$ leads to same $\alpha_c$ values beyond which the strong links play crucial role in the evolution of the model.   
\begin{figure}[ht]
\centering
\includegraphics[width=7cm]{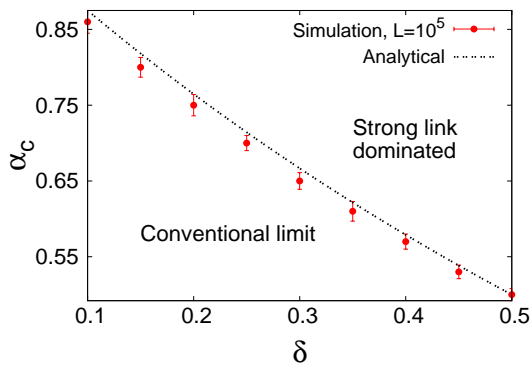}
\caption{Variation of $\alpha_c$ with strength of disorder in case of uniform threshold distribution. For $\alpha<\alpha_c$, the model operates in conventional limit. Beyond $\alpha_c$ the strong links play crutial role.}
\label{Alphac}
\end{figure}

Fig.\ref{Alphac} explicitly shows the variation of $\alpha_c$ with $\delta$. The value of $\alpha_c$ was observed at 0.5 earlier at $\delta=0.5$. The results on the present paper shows an increment in $\alpha_c$ as disorder strength is decreased. This suggest that as the disorder is decreased we have to go higher $\alpha$ values to make the model deviate from its conventional limit. 


\subsection{Stability and the high disorder limit}
We have already discussed the effect of low disorder on $\alpha_c$. In this section we want to focus on the question: what happens if the disorder is very high ? Understandably, at high disorder some links are already so strong that no additional strong link might be required at all. Such high disorder limit for the model is achieved by choosing a power law distribution (say, with power $-1$) for the threshold values instead of the uniform one. The threshold strength values are chosen between $10^{-\eta}$ and $10^{\eta}$; $\eta$ being the amount of disorder here.  

\begin{figure}[ht]
\centering
\includegraphics[width=7cm]{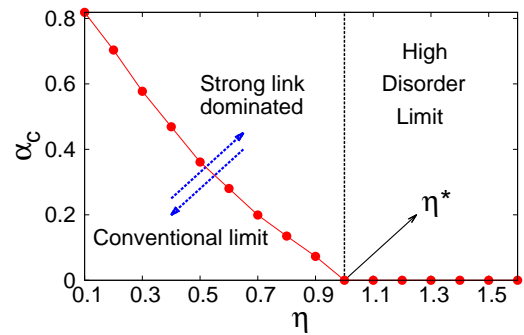}
\caption{$\alpha_c$ vs $\eta$ (strength of disorder) in case of power law threshold distribution. Beyond $\eta^{\ast}$, some fibers are itself so strong that no strong link requires to stabilize the model. Below $\eta^{\ast}$ we get two regions: conventional and strong link dominated, depending on whether $\alpha>\alpha_c$ or $\alpha<\alpha_c$.}
\label{Alphac_Pow}
\end{figure}

For power law distribution also, $\alpha_c$ shows a decreasing behavior with disorder $\eta$. The numerical result shows that we obtain $\alpha_c=0$ as we go beyond $\eta^{\ast}$. The region $\eta>\eta^{\ast}$ acts as the high disorder limit for the model, where the conventional burst statistics are not observed even in absence of any kind of strong links. 
\begin{itemize}
\item For $\eta>\eta^{\ast}$, we never obtain the conventional result for avalanche size distribution for the mean field fiber bundle model. The avalanche size distribution is not scale free in this limit. At a high $\eta$, the threshold values are so distinct to each other, we hardly observe any big avalanches.   
\item For $\eta<\eta^{\ast}$, the model either operates in the conventional limit, producing a scale free avalanche size distribution (with exponent -5/2) or dominated by the strong links, depending on whether $\alpha$ is less or greater than $\alpha_c$.  
\end{itemize}


\section{Discussions}
The present work deals with the existence of strong links in fiber bundle model. The model is considered to be equivalent to the composite material as it is a mixture of two types of fibers: $\alpha$ fraction unbreakable and $(1-\alpha)$ fraction with a certain threshold value. We observe that with increasing $\alpha$ not only failure abruptness decreases but also it increases the strength of the bundle. The brittle to quasi-brittle transition point scales to lower values when $\alpha$ is increased and that in turn reduces the brittle like response in the model. This in turn satisfies some criteria of being a composite. Also, the critical fraction of strong links ($\alpha_c$), beyond which the behavior of the model deviates from the conventional results, is observed to be a function of disorder strength $\delta$ or $\eta$. The change in behavior is captured beautifully through the sudden change in the average maximum avalanche during the failure process. In high disorder (high $\eta$ value) limit $\alpha_c$ keep decreasing and gradually reaches a trivial point $\alpha_c=0$ at $\eta=\eta^{\ast}$. This happens probably because at high $\eta$ the fibers are itself so strong that it does not need a fraction unbreakable to stabilize the failure process. The universality of the results are verified with three different threshold distribution: uniform, power law and Weibull.      


\section{Conclusion}
In conclusion, the recent study gives a clear idea how the failure process in the mean field limit is modified with variable strength of disorder, when the conventional fiber bundle model is combined with a fraction of infinitely strong fibers. Also critical fraction of strong fibers, required to deviate the model from the conventional results, scales down as we increase the strength of disorder. Due to the over simplicity of the model, one to one correspondence is not possible, though the results here can be compared loosely with the composite materials based on two basic observations : (i) increment in bundle strength and (ii) a decrease in failure abruptness.      
 

\section{Acknowledgement}
I thank Earthquake Research Institute, University of Tokyo, for the funding during this work. A special thank goes to Aakriti Saxena Roy for critical reading of the manuscript.


\end{document}